\documentstyle[procl,epsfig]{article}
\def\Journal#1#2#3#4{{#1} {\bf #2}, #3 (#4)}

\def\NPB{{\em Nucl. Phys.} B}

\def\PLB{{\em Phys. Lett.}  B}
\def\PLBo{{\em Phys. Lett.}}

\def\PRD{{\em Phys. Rev.} D}

\def\ZPC{{\em Z. Phys.} C}

\def\beq{\begin{equation}}

\def\eeq{\end{equation}}

\def\beqn{\begin{eqnarray}}

\def\eeqn{\end{eqnarray}}

\def\Proj{\Delta {\cal P}}

\def\as{\alpha_s}

\newcommand{\slsh}{\rlap{$\;\!\!\not$}}     % Feynman slash

\begin{document}

\begin{flushright}
\vspace*{-3.cm}
RAL-TR-96-046 \\
June 1996 \\ 
\vspace*{0.9cm}
\end{flushright}

%\title{THE POLARIZED TWO-LOOP SPLITTING FUNCTIONS}
\title{THE POLARIZED TWO-LOOP SPLITTING FUNCTIONS*}

\author{ W. VOGELSANG }

\address{Rutherford Appleton Laboratory,  
Chilton DIDCOT, Oxon OX11 0QX, England}

\maketitle\abstracts{
We present a brief description of the light-cone gauge calculation of the 
spin-dependent next-to-leading order splitting functions.}

It has recently become possible to perform analyses of the spin-dependent
parton distributions of a longitudinally polarized hadron at next-to-leading
order (NLO) accuracy of QCD. A first such phenomenological NLO study, taking
into account all available experimental data on polarized
deep-inelastic scattering has been presented in~\cite{grsv}, followed
by the analyses~\cite{mel}. An indispensable ingredient here are the polarized
two-loop splitting functions (or anomalous dimensions) $\Delta P^{(1)}_{ij}$ 
which appear in the NLO $Q^2$-evolution equations for the spin-dependent parton
densities. Results (in the $\overline{\mbox{MS}}$ scheme) for the $\Delta
P^{(1)}_{ij}$ have first been obtained in~\cite{MVN} where the Operator Product
Expansion (OPE) formalism was used. The results were afterwards confirmed
in~\cite{WV} using the somewhat more efficient method developed
in~\cite{EGMPR} and employed in the unpolarized case in~\cite{CFP,FP,EV},
which is based on the factorization properties of mass singularities and on 
the use of the axial gauge. In this paper we give a brief description of our 
calculation~\cite{WV}.

To begin with, we collect all ingredients for a NLO analysis of longitudinally 
polarized deep-inelastic scattering in terms of the spin-dependent structure 
function $g_1 (x,Q^2)$. Beyond LO, there are two different 
short-distance cross sections, $\Delta C_q$ and $\Delta C_g$, for scattering 
off incoming polarized quarks and gluons, respectively. 
Thus the NLO expression for $g_1$ reads in 
%general:
general:\footnote{* Invited talk presented at the
'Int. Workshop on Deep Inelastic Scattering and Related Phenomena' (DIS96),
Rome, Italy, April 15-19, 1996.}
\beqn
g_1 (x,Q^2) &=& \frac{1}{2} \sum_{i=1}^{n_f} e_i^2\; 
\Bigg\{ \Delta q_i (x,Q^2)+\Delta \bar{q}_i (x,Q^2)+ \nonumber \\
&+& \frac{\alpha_s(Q^2)}{2\pi} \left[ \Delta C_q \otimes 
\left( \Delta q_i +\Delta \bar{q}_i \right) +\frac{1}{n_f} 
\Delta C_g \otimes \Delta g\right] (x,Q^2) \Bigg\} \: ,  \label{g1}
\eeqn
where $n_f$ is the number of flavors and $\otimes$ denotes the usual 
convolution. Here, the polarized parton distributions $\Delta f \equiv
f^{\uparrow}-f^{\downarrow}$ ($f=q,\bar{q},g$) are to be evolved in 
$Q^2$ according to the NLO spin-dependent Altarelli-Parisi~\cite{ap} 
evolution equations. We adopt the following perturbative expansion of 
the evolution kernels:
\beq \label{expan}
\Delta P_{ij} (x,\as) = \left( \frac{\as}{2\pi} \right) \Delta 
P_{ij}^{(0)} (x) + \left( \frac{\as}{2\pi} \right)^2 
\Delta P_{ij}^{(1)} (x) + \ldots   \; .
\eeq
We emphasize that neither of the NLO corrections, $\Delta C_i$ and 
$\Delta P_{ij}^{(1)}$, are physical quantities since they depend on the 
factorization scheme adopted. Needless to mention that the same scheme has 
to be chosen in the calculation of both in order to obtain a meaningful 
result. Conversely, once the $\Delta C_i$ and $\Delta P_{ij}^{(1)}$
are known in one scheme it is possible to perform a factorization scheme 
transformation, i.e., to shift terms between them without changing a physical 
quantity like $g_1$, hereby redefining the polarized NLO parton 
distributions~\cite{GR}. 
 
Defining $\Delta q^\pm_i \equiv \Delta q_i \pm \Delta \bar{q}_i$
one finds the following NLO evolution equations for the non-singlets (NS)
$\Delta q_i^-$ and $\Delta q_i^+ - \Delta q_j^+$: 
\beqn
\frac{d}{d\ln Q^2} (\Delta q_i^+ - \Delta q_j^+) &=& 
\Delta P_{qq}^+ (x,\as (Q^2)) \otimes (\Delta q_i^+ - \Delta q_j^+) 
\; , \label{a3evol} \\
\frac{d}{d\ln Q^2} \Delta q_i^- &=& 
\Delta P_{qq}^- (x,\as (Q^2)) \otimes \Delta q_i^- \; ,  
\eeqn
where we have suppressed the obvious argument $(x,Q^2)$ in all
parton densities and taken into account that there are two different
NS splitting functions, $\Delta P_{qq}^{\pm}$, beyond LO (see,
e.g.,~\cite{EV}). Defining $\Delta \Sigma \equiv \sum_i (\Delta q_i+\Delta 
\bar{q}_i)$ one has in the flavor singlet sector:
\begin{eqnarray} \label{AP}
\frac{d }{d \ln Q^2}   
\left( \begin{array}{c} \Delta \Sigma  \\ \Delta g 
\end{array} \right)
= \left( \begin{array}{cc} \Delta P_{qq}(x,\as (Q^2)) &  
\Delta P_{qg}(x,\as (Q^2))\\  
\Delta P_{gq}(x,\as (Q^2) ) &  \Delta P_{gg}(x,\as (Q^2) )\\  
\end{array}\right)  \otimes
\left(  \begin{array}{c}
\Delta \Sigma  \\ 
\Delta g
\end{array} \right) \: . \hspace*{-6pt}
\end{eqnarray}
The $qq$-entry in the singlet matrix of splitting functions is written as
\beq  \label{qqs}
\Delta P_{qq}=\Delta P_{qq}^+ + \Delta P^S_{qq} \; .
\eeq
Thus, at NLO, we will have to derive the splitting functions $\Delta 
P_{qq}^{\pm ,(1)}$, $\Delta P_{qq}^{S, (1)}$, and those involving gluons.
The general strategy to do this in the method of~\cite{EGMPR,CFP,FP} consists 
of first expanding the squared matrix element $\Delta M$ for (polarized) 
virtual photon--polarized  quark (gluon) scattering into a ladder of 
two-particle irreducible (2PI) kernels~\cite{EGMPR} $C_0$, $K_0$,
\beq \Delta M = \Delta \Bigg[ C_0
(1+K_0+K_0^2+K_0^3+\ldots) \Bigg] \equiv \Delta \Bigg[ \frac{C_0}{1-K_0} 
\Bigg] \; .
\eeq 
We now choose the light-cone gauge by introducing a light-like 
vector $n$ ($n^2=0$) with $n \cdot A=0$. In this  gauge the 2PI kernels are 
finite as long as external legs are kept unintegrated, such that collinear 
singularities only appear when integrating over the lines connecting 
the rungs of the ladder~\cite{EGMPR}. This allows for projecting out the 
singularities by introducing the projector onto polarized physical states, 
$\Proj$. Thus $\Delta M$ can be written in the factorized form 
\beq
\Delta M = \Delta C \Delta \Gamma \;,
\eeq
where $\Delta C = \Delta C_0/(1-(1-\Proj )K_0)$ is the
(finite) short-distance cross section, whereas $\Delta \Gamma$ contains
all (and only) mass singularities. Working in dimensional regularization
($d=4-2 \epsilon$) in the $\overline{\mbox{MS}}$ scheme one has
explicitly~\cite{CFP}:
\beq \label{gam}
\Delta \Gamma_{ij} = Z_{j} \Bigg[\delta(1-x) 
\delta_{ij}+x \; \mbox{PP} \int \frac{d^dk}{(2 \pi)^d}
\delta(x-\frac{kn}{pn}) \Delta U_i K \frac{1}{1-\Proj K} 
\Delta L_j\Bigg] \: , 
\eeq 
where `PP' extracts the pole part, $Z_j$ ($j=q(g)$) is the residue of the 
pole of the full quark (gluon) propagator, and we have defined 
$K=K_0 (1 - (1-\Proj) K_0)^{-1}$. $k$ is the momentum of the parton 
leaving the uppermost kernel in $\Delta \Gamma$. The spin-dependent 
projection operators onto physical states are given by
\beq  \label{ulproj}
\Delta U_q =-\frac{1}{4 kn} \gamma_5 \slsh{n},\;\;
\Delta L_q = -\slsh{p} \gamma_5 \; ; \;\; 
\Delta U_g = i \epsilon^{\mu\nu\rho\sigma} \frac{n_{\rho} k_{\sigma}}
{kn} , \;\; \Delta L_g = i \epsilon^{\mu\nu\rho\sigma} \frac{p_{\rho} 
n_{\sigma}} {2pn} \; .
\eeq
Finally, it can be shown~\cite{CFP} that the coefficient of the 
$1/\epsilon$ pole of $\Delta \Gamma$ is related to the splitting functions 
we are looking for:
\beq
\Delta \Gamma_{qq} = \delta (1-x) - 
\frac{1}{\epsilon} \Bigg(\frac{\as}{2\pi} \Delta P_{qq}^{(0)}(x)+\frac{1}{2} 
\left( \frac{\as}{2\pi} \right)^2 \Delta P_{qq}^{(1)} (x) + \ldots \Bigg) + 
O \left(\frac{1}{\epsilon^2} \right) 
\eeq
and analogously for the flavor singlet case. Explicit examples of the Feynman 
diagrams contributing to the $\Delta \Gamma_{ij}$ can be found 
in~\cite{CFP,WV,EV}.

We see from Eq.~(\ref{ulproj}) that there is a new ingredient in the
polarized calculation which requires extra attention: The Dirac matrix 
$\gamma_5$ and the Levi-Civita tensor 
$\epsilon_{\mu\nu\rho\sigma}$ enter. A prescription for dealing with these 
(genuinely {\em four}-dimensional) quantities in $d=4-2 \epsilon$ dimensions 
has to be adopted which must be free of algebraic inconsistencies.
Our calculation~\cite{WV} was performed using the original
definitions for $\gamma_5$ and $\epsilon_{\mu\nu\rho\sigma}$ of~\cite{HVBM}
(HVBM scheme) which is usually regarded as the most reliable prescription.
Here $\gamma_5$ retains its {\em four}-dimensional definition, 
$\gamma_5 \equiv i \epsilon^{\mu\nu\rho\sigma} \gamma_{\mu}
\gamma_{\nu}\gamma_{\rho}\gamma_{\sigma}/4!$, with the $\epsilon$-tensor 
being a genuinely four-dimensional object. As a consequence one 
finds that
\beq
\left\{ \gamma^{\mu},\gamma_5 \right\} = 0  \;\;\; \mbox{for} 
\; \mu=0,1,2,3 \; ;  \;\;\; 
\left[ \gamma^{\mu},\gamma_5 \right] = 0  \;\;\; \mbox{otherwise}  \; .
\label{g5anti}
\eeq
Thus the matrix element squared of a graph will in general depend on scalar 
products defined in the '$(d-4)$-dimensional' subspace. 
Special care has to be taken of such terms in loop and phase space 
integrals~\cite{WV}.

Another more technical remark concerns the use of the light-cone gauge,
which plays a crucial role in the calculation. The light-cone gauge 
denominator $1/(n\cdot l)$ in the gluon propagator can give rise to 
additional divergencies in loop and phase space integrals. 
We follow~\cite{CFP,FP,EV} to use the principal value (PV) prescription 
to regulate such poles:
\beq \label{PPprescription}
\frac{1}{n\cdot l} \rightarrow \frac{1}{2} \Bigg(
\frac{1}{n\cdot l+ i \delta (pn)} + 
\frac{1}{n\cdot l- i \delta (pn)} \Bigg)
=\frac{n \cdot l}{(n\cdot l)^2 + \delta^2 (pn)^2} \; .
\eeq 
The  PV prescription appears to be the most convenient choice from a 
practical point of view; it leads, however, to the feature that the 
renormalization 'constants' depend~\cite{CFP,EV} on the longitudinal momentum 
fractions $x$.

We express the $\overline{\mbox{MS}}$ results of our calculation in the 
HVBM scheme in terms of the unpolarized NLO NS splitting functions 
$P_{qq}^{\pm,(1)}$ of~\cite{CFP} and of the recent polarized OPE 
results $\Delta 
\tilde{P}_{ij}^{(1)}$ of~\cite{MVN}, exploiting the fact that 
the contributions $\sim \delta(1-x)$ to the diagonal splitting 
functions are necessarily the same as in the unpolarized case
since they are determined by $Z_j$ in (\ref{gam}). One then has:
\begin{eqnarray} \label{inter}
\Delta P_{qq}^{\pm ,(1)} (x) &=& P_{qq}^{\mp ,(1)} (x) - 2 \beta_0 
C_F (1-x) \:\:\: , \nonumber  \\
\Delta \hat{P}^{(1)} (x) &=& \Delta \hat{\tilde{P}}^{(1)}
(x) -\frac{\beta_0}{2} \hat{A}(x) + \left[\hat{A}(x),\hat{P}^{(0)} (x) 
\right]_{\otimes} \;\; , \nonumber  \\
\Delta C_q (x) &=& \Delta \tilde{C}_q (x)
-4 C_F (1-x) \:\:\: , \nonumber \\
\Delta C_g (x) &=& \Delta \tilde{C}_g (x) \:\:\: ,
\end{eqnarray}
where $\hat{P}^{(0)}$ and $\hat{P}^{(1)}$ denote the LO and NLO parts,
respectively, of the singlet evolution matrix, and 
$$ \hat{A} (x) \equiv 4 C_F (1-x) \left( \begin{array}{cc}
                                  1 & 0 \\
                                  0 & 0
                                  \end{array} \right) \;\; . $$
In Eq.~(\ref{inter}) we have also included the results for the 
short-distance cross sections $\Delta C_q$, $\Delta C_g$. As indicated in 
Eq.~(\ref{inter}), the '$+$' and '$-$' combinations of the NS splitting 
functions have interchanged~\cite{grsv,ms} their roles. 
Eqs.~(\ref{a3evol},\ref{inter}) therefore 
imply that the combination $\Delta P_{qq}^{+,(1)}=P_{qq}^{-,(1)} - 
2\beta_0 C_F (1-x)$ would govern the $Q^2$-evolution of, e.g., the 
polarized NS quark combination 
$$\Delta A_3 (x,Q^2) = \left( \Delta u^+ - \Delta d^+ \right) (x,Q^2)
\:\:\: .$$
Since the first moment (i.e., the $x$-integral) of the latter corresponds
to the nucleon matrix element of the NS axial vector current
$\bar{q} \gamma^{\mu} \gamma_5 \lambda_3 q$ which is conserved, 
it has to be $Q^2$-independent. Keeping in mind that 
the integral of the unpolarized $P_{qq}^{-,(1)}$ vanishes already due to 
fermion number conservation~\cite{CFP}, it becomes obvious that the additional 
term $-2 \beta_0 C_F (1-x)$ in (\ref{inter}) spoils the $Q^2$-independence 
of the first moment of $\Delta A_3 (x,Q^2)$. On the other hand, as pointed
out earlier, we are free to perform a factorization scheme transformation 
to the results in (\ref{inter}). It turns out~\cite{WV} that the scheme 
transformation that removes the term $-2 \beta_0 C_F (1-x)$ from 
$\Delta P_{qq}^{\pm ,(1)}$ in Eq.~(\ref{inter}) eliminates {\em at the same 
time all} extra $(1-x)$-terms on the r.h.s. of (\ref{inter}), leaving 
$\Delta C_g$ unchanged. Thus our final results after the transformation 
are in complete agreement with those of~\cite{MVN}. 
We note that the presence of the $(1-x)$-terms 
in our original HVBM scheme result (\ref{inter}) can be traced
back to the fact that in this scheme the $d$-dimensional polarized LO 
quark-to-quark splitting function is no longer equal to its unpolarized 
counterpart due to the non-anticommutativity of $\gamma_5$ 
(see (\ref{g5anti})), artificially violating helicity conservation.

\begin{figure}[htb]
\vspace*{-0.8cm}
\hspace*{-0.2cm}
\epsfig{file=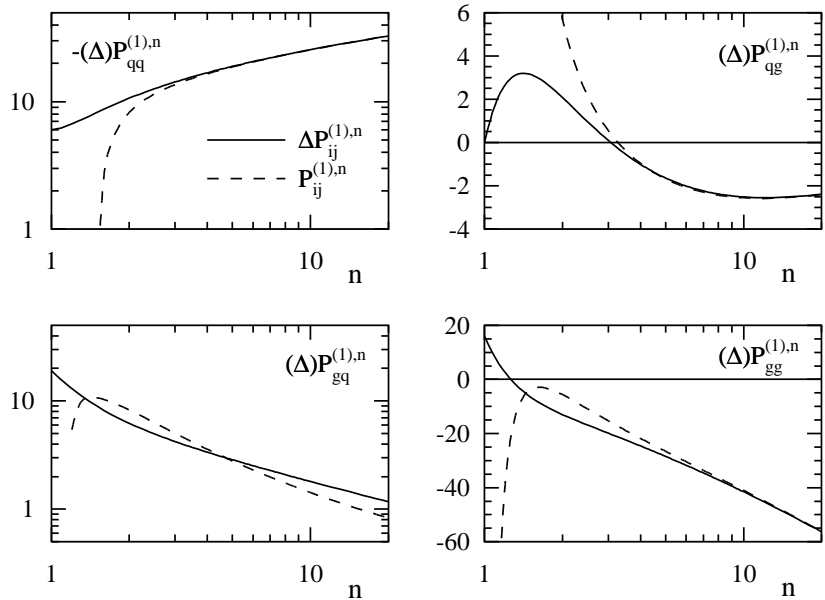,width=12.8cm,angle=0}
\vspace*{-1cm}
\caption{Mellin-moments of the polarized and unpolarized
NLO singlet splitting functions.}
\label{split}
\vspace*{-0.2cm}
\end{figure}

Our complete final results for the $\Delta P_{ij}^{(1)} (x)$ can be found 
in~\cite{WV} and need not be repeated here. We mention that compact 
expressions for the Mellin-moments of the polarized NLO splitting functions,
defined by 
\beq 
\Delta P_{ij}^{(1),n} \equiv \int_0^1 x^{n-1} \Delta P_{ij}^{(1)} (x) dx
\eeq
as well as their analytic continuations to arbitrary complex $n$, can be 
found in~\cite{grsv}. To work in Mellin-$n$ space is very convenient for a 
numerical analysis of parton distributions since the evolution 
equations can be solved analytically here. For illustration we show the entries 
$\Delta P_{ij}^{(1),n}$ of the NLO part of the singlet evolution matrix
as a function of real Mellin-$n$ in Fig.~\ref{split}, comparing 
them to the unpolarized 
$P_{ij}^{(1),n}$ as obtained from~\cite{flor}. One observes, in particular, 
that $\Delta P_{ij}^{(1),n} \rightarrow P_{ij}^{(1),n}$ for $n\rightarrow 
\infty$ (i.e., for $x\rightarrow 1$), except~\cite{grsv} for 
$\Delta P_{gq}^{(1),n}$.
We finally note that the values for the first moments of the 
$\Delta P_{ij}^{(1)}(x)$ turn out to be~\cite{MVN,WV,kod}
\beqn
&&\Delta P_{qq}^{(1),n=1} = -3 C_F T_f  \;\; , \;\;\;     
  \Delta P_{gq}^{(1),n=1} = -\frac{9}{4} C_F^2 + 
\frac{71}{12} N_C C_F - \frac{1}{3} C_F T_f \; , \nonumber \\ 
&&\Delta P_{qg}^{(1),n=1} = 0  \;\; , \;\;\; 
  \Delta P_{gg}^{(1),n=1} = \frac{17}{6} N_C^2 - C_F T_f
-\frac{5}{3} N_C T_f \equiv \frac{\beta_1}{4} \; ,
\eeqn
where $C_F=4/3$, $N_C=3$, $T_f=n_f/2$. 
 

\section*{References}
\begin{thebibliography}{99}
\bibitem{grsv} M.\ Gl\"{u}ck, E.\ Reya, M.\ Stratmann,
W.\ Vogelsang,  \Journal{\PRD}{53}{4775}{1996}.   
\bibitem{mel} R.D.\ Ball, S.\ Forte, G.\ Ridolfi,
\Journal{\PLB}{378}{255}{1996}; \\
T.K.\ Gehrmann, W.J.\ Stirling, \Journal{\PRD}{53}{6100}{1996}.
\bibitem{MVN} R.\ Mertig, W.L.\ van Neerven, \Journal{\ZPC}{70}{637}{1996}.
\bibitem{WV} W.\ Vogelsang, Rutherford RAL-TR-95-071, to appear in 
{\em Phys. Rev.} D; RAL-TR-96-020, to appear in {\em Nucl. Phys.} B.
\bibitem{EGMPR} R.K. Ellis {\it et al.}, \Journal{\NPB}{152}{285}{1979}. 
\bibitem{CFP} G. Curci, W. Furmanski, R. Petronzio, 
\Journal{\NPB}{175}{27}{1980}. 
\bibitem{FP} W. Furmanski, R. Petronzio, 
\Journal{\PLBo}{97B}{437}{1980}.
\bibitem{EV} For a detailed description of the calculation in the unpolarized
case see: R.K. Ellis, W. Vogelsang, RAL-TR-96-012, hep-ph/9602356.
\bibitem{ap} G. Altarelli, G. Parisi, \Journal{\NPB}{126}{298}{1977}.       
\bibitem{GR} see, e.g., M. Gl\"{u}ck, E. Reya, \Journal{\PRD}{25}{1211}{1982}.
\bibitem{HVBM} G. 't Hooft, M. Veltman, \Journal{\NPB}{44}{189}{1972}; \\
P. Breitenlohner, D. Maison, {\em Comm. Math. Phys.} {\bf 52}, 11 (1977).
\bibitem{ms} M. Stratmann, W. Vogelsang, A. Weber, 
\Journal{\PRD}{53}{138}{1996}.
\bibitem{flor} E.G. Floratos, C. Kounnas, R. Lacaze, 
\Journal{\NPB}{192}{417}{1981}.
\bibitem{kod} J. Kodaira, \Journal{\NPB}{165}{129}{1980}; 
G. Altarelli, B. Lampe, \Journal{\ZPC}{47}{315}{1990}; 
S.A. Larin, \Journal{\PLB}{303}{113}{1993}. 
\end{thebibliography}
\end{document}